\documentclass[prl,twocolumn,amsmath,amssymb]{revtex4}

\usepackage{graphicx}
\usepackage{dcolumn}
\usepackage{bm}
\usepackage{mathrsfs}
\usepackage{appendix}
\usepackage{bbm}
\usepackage{color}

\def\be{\begin{equation}}
\def\ee{\end{equation}}
\def \uudd {\uparrow \uparrow \downarrow \downarrow}
\def\bd{\begin{displaymath}}
\def\ed{\end{displaymath}}
\def\-{\phantom{-}}

\begin{document}

\title{Identifying the multiferroic phase of doped CuFeO$_2$ using inelastic neutron scattering}

\author{J.T. Haraldsen$^1$, F. Ye$^2$, R.S. Fishman$^1$, J.A. Fernandez-Baca$^{2,3}$, Y. Yamaguchi$^4$,
K. Kimura$^4$ and T. Kimura$^4$}

\affiliation{$^1$Materials Science and Technology Division, Oak Ridge National Laboratory, \mbox{Oak Ridge, TN 37831}}
\affiliation{$^2$Neutron Scattering Science Division, Oak Ridge National Laboratory, \mbox{Oak Ridge, TN 37831}}
\affiliation{$^3$Department of Physics and Astronomy, The University of Tennessee, \mbox{Knoxville, TN 37996}}
\affiliation{$^4$Division of Materials Physics, Graduate School of Engineering Science, Osaka University, Osaka, \mbox{560-8531, Japan}}

\pacs{78.70.Nx, 75.30.Ds, 75.10.Jm,75.30.Et}

\begin{abstract}

 We report inelastic neutron scattering measurements that provide a distinct dynamical ``fingerprint" for the multiferroic ground state of 3.5\% Ga-doped CuFeO$_2$. The complex ground state is stabilized by the displacement of the oxygen atoms, which are also responsible for the multiferroic coupling predicted by Arima. By comparing the observed and calculated spectrum of spin excitations, we conclude that the magnetic ground state is a distorted screw-type spin configuration that requires a mechanism for magnetoelectric coupling different from the generally accepted spin-current model.

\end{abstract}

\maketitle

Multiferroic materials allow the electric polarization to be controlled by
switching the direction of magnetic ordering and consequently offer
prospects for many new technological applications  \cite{kimu:03,khom:06,cheo:06,rame:07,Eere:06}. 
Because multiferroic behavior has been found in materials that exhibit complex
(non-collinear and incommensurate) magnetic order, determining the spin arrangement of
the ground states is essential to understand its role in the multiferroic coupling 
\cite{kats:05,Eere:06,most:06,serg:06,arim:07,soda:09}. 
In the spin-current model or inverse Dzyaloshinskii-Moriya mechanism
of multiferroelectricity \cite{most:06,kats:05,serg:06}, the electric polarization is induced by
inversion symmetry breaking associated with a complex spin spiral structure.
This model predicts that the electric polarization is perpendicular to both
the chiral axis ${\bf S}_i \times {\bf S}_j$ and the magnetic ordering wavevector {\bf Q}. However, in
materials like CuFeO$_2$, the electric polarization is parallel \cite{naka:07,seki:07,kane:07} to {\bf Q},
indicating that the spin-current model cannot explain the multiferroic
coupling. The multiferroic behavior of CuFeO$_2$ can instead be explained \cite{arim:07}
by spin-orbit coupling in the presence of lattice distortions that lower the
crystal symmetry. Therefore, it is important to understand how lattice
distortions affect the magnetic structure of doped CuFeO$_2$.

Nakajima $et~al.$ \cite{naka:07} proposed that the magnetic structure of multiferroic
CuFeO$_2$ is a ``proper" spiral configuration composed of two alternating simple
spiral structures.  Fishman and Okamoto \cite{fish:10}, on the other hand, reported
that the ground state for a frustrated triangular lattice has a complex
non-collinear (CNC) spin configuration characterized by the presence of
higher spin harmonics. Elastic neutron scattering measurements alone are not
sufficient to distinguish between these two complex magnetic states because
they produce similar sets of observable satellite reflections.
Consequently, novel ways must be devised to determine the true magnetic
ground state. This Letter reports inelastic neutron scattering (INS)
measurements that provide a distinct dynamical ``fingerprint" for the
multiferroic ground state of 3.5\% Ga-doped CuFeO$_2$. From our analysis, we
conclude that the magnetic ground state is a distorted screw-type spin
configuration that is incompatible with the generally accepted spin-current
model. This complex ground state is stabilized by the displacement of the
oxygen atoms, which are also responsible for the multiferroic coupling
predicted by Arima \cite{arim:07}.

\begin{figure}
\includegraphics[width=2.8in]{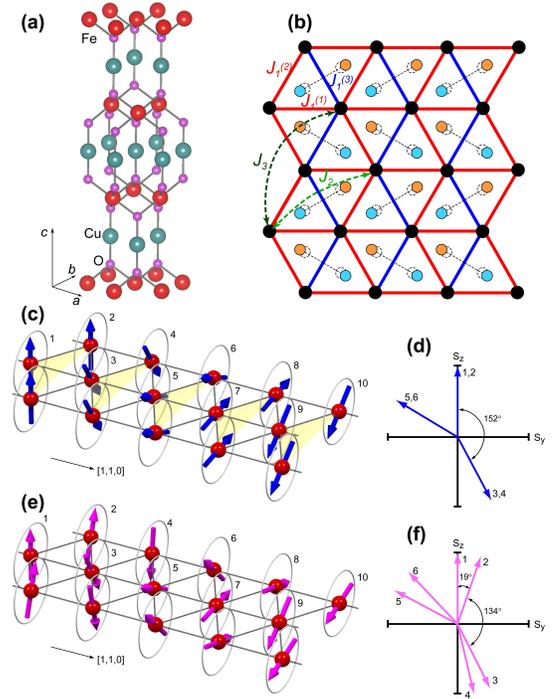}
\caption{(a) The three-dimensional crystal structure of CuFeO$_2$. (b) The two-dimensional hexagonal lattice structure of the Fe$^{3+}$ layers and the intralayer interactions. The orange and blue colored atoms illustrate the displacements of oxygen atoms lying above and below the hexagonal plane. The magnetic structure and $yz$ projections of the ``proper" (c) and  (d) and complex non-collinear (CNC) (e) and (f) spiral or screw-type spin configurations are also shown.}
\label{spinconfig}
\end{figure}

\begin{figure}
\includegraphics[width=3.25in]{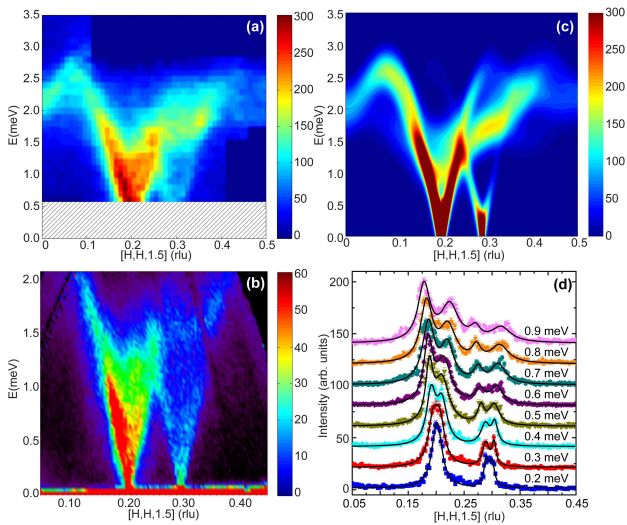}
\caption{Energy versus momentum transfer taken from the triple-axis spectrometer at  HFIR (a) and the  time-of-flight Cold Neutron Chopper Spectrometer (CNCS) at the SNS (b). (c) Predicted simulation of the INS data using intra- and interlayer interactions along with a lattice distortion and anisotropy. (d) Wavevector scans with different energy transfers for the CNCS data.}
\label{fig2}
\end{figure}

With a large $S$ = 5/2 magnetic moment contributed by the Fe$^{3+}$ ions, CuFeO$_2$ (Fig.~\ref{spinconfig}(a)) has inspired great interest due to the magnetic frustration within each hexagonal plane \cite{ye:07}, which results in a four sublattice collinear spin configuration characterized by the propagation wavevector [0.25,0.25,1.5].  A multiferroic state can be induced in this system by applying a magnetic field above about 7 T  \cite{kimu:06} or by doping with non-magnetic Al or Ga impurities \cite{seki:07,kane:07}. The resulting multiferroic ground state is non-collinear and characterized by the incommensurate propagation wavevectors Q = [H,H,1.5] where H $\approx$ 0.20 and 0.30 \cite{tera:05,tera:08}. It is notable that the spin excitation spectrum of the collinear non-multiferroic phase of pure CuFeO$_2$ softens at the incommensurate wavevectors characteristic of the multiferroic phase \cite{ye:07}. This softening can be simulated by lowering the anisotropy energy \cite{fish:08jap} and the complete softening of this mode signals the development of the multiferroic non-collinear spin structure \cite{swan:09,hara:09prl}. The observed softening of the spin excitations of CuFeO$_2$ can be interpreted as a dynamical precursor of the multiferroic phase.

Determining the spin configuration of multiferroic CuFeO$_2$ is nontrivial. A simple spiral structure, in which the average turn angle of Fe spins is 74$^{\circ}$, would produce the observed peak at [0.20,0.20,1.5] but not the peak at [0.30,0.30,1.5] . Nakajima $et~al.$ \cite{naka:07} proposed a ``proper'' spiral configuration (Fig.~\ref{spinconfig}(c) and (d)) with two alternating turn angles of 0$^{\circ}$ and 152$^{\circ}$ that will produce the two main observed peaks [H,H,1.5] (H $\approx$ 0.20 and 0.30).

Recently, several groups have observed lattices distortions associated with displacements of the oxygen atoms (illustrated in Fig.~\ref{spinconfig}(b)) \cite{ye:06,tera:06a}.  Incorporating these distortions, Fishman and Okamoto found that odd-order spin harmonics produce a magnetic ground state with a CNC spin configuration (Fig. \ref{spinconfig}(e) and (f)) \cite{fish:10}. The oxygen displacements produce a modulation in the nearest neighbor interactions $J_1$ (Fig.~\ref{spinconfig}(b)) that breaks the equilateral symmetry (Fig. \ref{spinconfig}(b)) of the lattice with nearest-neighbor exchange interactions $J_1^{(1)}$ = $J_1^{(2)}$ = $J_1 - K_1$/2 and $J_1^{(3)}$ = $J_1 + K_1$, where $K_1$ is a measure of the atomic distortion.  The distortion creates turn angles that fluctuate around 19$^{\circ}$ and 134$^{\circ}$. For a distorted antiferromagnet, a CNC screw-type configuration is energetically favored over both simple and ``proper" spirals \cite{fish:10}. Elastic neutron scattering measurements cannot distinguish between the ``proper" spiral and CNC configurations. 

The hexagonal lattice symmetry of CuFeO$_2$ provides a complex network of multiple intra- and interlayer super-exchange pathways (Fig.~\ref{spinconfig}(b)) \cite{fish:08}.  The Heisenberg Hamiltonian
can be written as
\be
H = -\frac{1}{2}\sum_{i \neq j} J_{ij} \mathbf{{S}}_i \cdot \mathbf{{S}}_j - D \sum_i \mathbf{{S}}_{iz}^2,
\label{genH}
\ee
where $\mathbf{S}_i$ is the local moment on site $i$, $D$ is the anisotropy energy, and the exchange coupling $J_{ij}$ between sites $i$ and $j$ is antiferromagnetic when $J_{ij}  < 0$.  

To incorporate the spin harmonics, we modified the classical approach described in Ref. [\onlinecite{fish:10}] by defining $S_z$ within any hexagonal plane as
\be
\begin{array}{c}
\displaystyle S_z(\mathbf{R})  = A \Big\{\cos(Q\cdot x) + \sum_{l=1}C_{2l+1}\cos[Q(2l+1)\cdot x] \\ \displaystyle - \sum_{l=0}B_{2l+1}\sin[(2\pi-Q)(2l+1)\cdot x ] \Big\},
\end{array}
\ee
where the $C_{2l+1}$ harmonics are produced by the anisotropy energy $D$ and the $B_{2l+1}$ harmonics are produced by the lattice distortion $K_1$.  The square of these harmonics 
are proportional to the observed elastic intensities at odd multiples of $Q$ and $2\pi -Q$.  The function $S_z({\bf R})$ is normalized so that the maximum of $|S_z(\mathbf{R})|$ is $S$ = 5/2.  
The perpendicular spin $S_y$ is given by
\be
\displaystyle S_y(\mathbf{R})  = \sqrt{S-S_z(\mathbf{R})^2} \, \rm{sgn}(g(\mathbf{R})), \ee 
where
\be
g(\mathbf{R}) = \sin(Q\cdot x)+G_1\cos[(2\pi-Q)\cdot x ]
\ee
and $G_1$ is an additional variational parameter.  

Hexagonal spin planes were then stacked along the $c$ axis and coupled by the exchange interactions $J_{zn}$.
The three-dimensional spin configuration was obtained by minimizing the energy on a large unit cell with length 10$^4$.  The spin fluctuations about this
three-dimensional spin configuration were evaluated using the method 
described in Ref. [\onlinecite{hara:09jcp}], where the frequencies and intensities of the SW excitations are evaluated simultaneously.   
We emphasize that a stable ground state is required to evaluate the spin dynamics:  a $1/S$ spin-wave expansion cannot be performed 
starting from the ``proper" helix sketched in Fig. \ref{spinconfig}(c).

\begin{figure}
\includegraphics[width=3.1in]{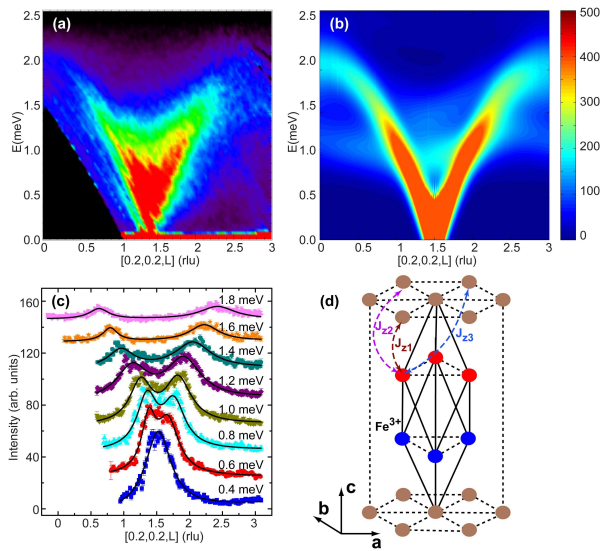}
\caption{(a) Energy versus momentum transfer taken from the CNCS clearly showing the SW modes along the [0,0,L] direction with H = 0.2. The 
dispersion with L indicates the importance of interlayer interactions.  (b) Predicted INS spectra using intra- and interlayer interactions along with a lattice distortion. (c) Intensity versus momentum transfer for various energies. (d) A three dimensional view of the Fe$^{3+}$ lattice structure illustrating the interlayer interactions. }
\label{fig3}
\end{figure}

Single crystals of CuFe$_{0.965}$Ga$_{0.035}$O$_2$ were grown using the floating zone technique. The crystals were oriented with two orthogonal wavevectors of [1,1,0] and [0,0,3] aligning in the horizontal plane. The INS measurements were carried out using the Cold Neutron Chopper Spectrometer (CNCS) spectrometer at the Spallation Neutron Source (SNS) and the HB-1 triple-axis spectrometers at the High Flux Isotope Reactor (HFIR) at the Oak Ridge National Laboratory. The low energy excitations were measured at 1.4 K with final energy chosen at 5.1 meV at HB-1 and with incident neutron energy chosen at 3 meV to ensure high resolution at CNCS.

\begin{figure}
\includegraphics[width=3.25in]{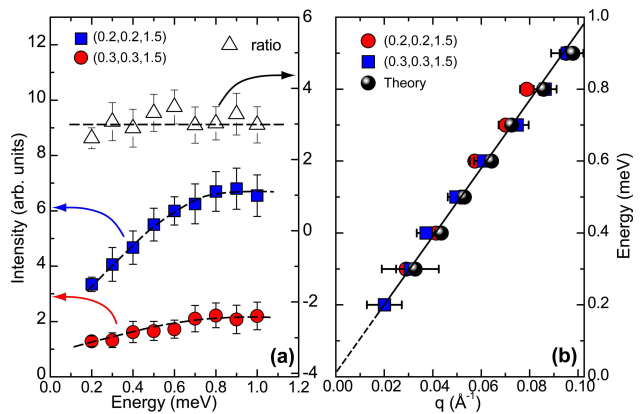}
\caption{(a) Intensity versus energy for the [0.20,0.20,1.5] (blue squares) and [0.30,0.30,1.5] (red circles) modes along the [H,H,0] direction with L = 1.5. The ratio of mode intensities shows that the main [0.20,0.20,1.5] mode is three times as intense as the [0.30,0.30,1.5] mode. The simulated intensity is consistent with this intensity ratio. (b) SW velocities for the two modes along the [H,H,0] direction with L = 1.5 together with the predicted values (gray circles).}
\label{fig4}
\end{figure}

Figures \ref{fig2}(a) and (b) present the INS data for the multiferroic phase of 3.5$\%$ Ga doped CuFeO$_2$ along [H,H,0] with L = 1.5.  ``Gapless" excitations are observed at H $\approx$ 0.20 and 0.30 together with a ``shoulder" at H $\approx$ 0.08 and an intensity ``hole" around H $\approx$ 0.30 and E $\approx$ 1.0 meV. Figure \ref{fig2}(c) displays the the predicted INS spectra \cite{interactions}, which contains contributions from both the normal spin configuration with wavevector along [H,H,1.5] and the two twins with
wavevectors along [H,0,1.5] and [0,H,1.5].  Strongly affected by the lattice distortion, the twin states account for the spectral weight around E $\approx$ 1.5 meV and H $\approx$ 0.30.  The model also accurately produces the intensity ``hole'' at 
H $\approx$ 0.30 as well as the ``shoulder" at H $\approx$ 0.08, which is caused by the interlayer interactions.  With the number of intricate features reproduced, we are confident that the dynamical ``fingerprint" of doped CuFeO$_2$ has been matched.

To compare the experimental and theoretical results more closely, we examined the SW velocities $d\omega$/$dq$ at low energy (Fig.~\ref{fig4}(b)).  The data points are determined by the energy cuts plotted in Fig.~\ref{fig2}(d).  Since the SW velocities are functions of the exchange interactions, this comparison assures us that the correct exchange parameters have been used. The exchange parameters $J_n$ and $J_{zn}$ are quite close \cite{interactions} to those predicted in Ref. [\onlinecite{fish:08}] from fits to the SW spectra of the $\uudd $ phase.  The experimental results show no evidence of an energy gap present in the multiferroic phase within the experimental energy resolution. This is consistent with our model, in which rotations of the spin plane about the $c$ axis cost no energy and a SW gap is absent.  However, the magnetostrictive energy that produces the observed confinement \cite{naka:07} of the spins to the [1,1,0] plane may induce a small gap in the excitation spectrum.

While four harmonics are predicted, only the $B_1$ component produces sufficient scattering intensity comparing to the main mode at [0.20,0.20,1.5] to be seen by our INS measurements \cite{interactions}. While the main SW mode is associated with the ordering wavevector $Q$, the secondary mode centering at [0.30,0.30,1.5] is produced by the $B_1$ harmonic at $2\pi -Q$. The overall magnitude of these modes are in agreement with our calculated values. For lower Ga doping or larger anisotropy, the higher harmonics will become more important and the distribution of turn angles will become more uniformly distributed. If $D$ is increased from 0.01 to 0.04 meV (the critical value above which the $\uudd $ phase is stable), then the amplitudes $C_3$ and $B_3$ rise by factors of 2 and 4, respectively.

Similar to the collinear excitation observed \cite{ye:07} in pure CuFeO$_2$, the dispersive excitation along the [0,0,L] direction through the magnetic Bragg peak at H $\approx$ 0.20 contains a minimum at L = 1.5 (Figs.~\ref{fig3}(a) and (b)). This underscores the importance of the interlayer interactions $J_{zn}$ (Fig. \ref{fig3}(d)) and the three dimensional character of this system. In comparison to pure CuFeO$_2$, the magnetic interactions $J_{z2}$ and $J_{z3}$ are somewhat weakened \cite{interactions}, which may be attributed to the disorder caused by Ga doping. Figure \ref{fig3}(c) shows the experimental cuts in energy demonstrating the splitting of the Goldstone mode at L $\approx$ 1.5.

Figure \ref{fig4}(a) compares the integrated scattering intensity originating from the main mode at H = 0.20 and the secondary mode at H = 0.30. The main mode is about three times as intense as the secondary one.   The simulation with $B_1$ = -0.52 yields an intensity ratio of 3.7, in good agreement with the measured value. The only well-defined modes in our model are the Goldstone modes emerging from [0.20,0.20,1.5] and [0.30,0.30,1.5].  
A continuum of excitations borders the Goldstone modes due to the complex and incommensurate nature of the magnetic structure. This contrasts with the relatively large gapped excitation and associated magnon-phonon hybridized electro-magnon observed in TbMnO$_3$ \cite{senf:07}.

According to Ref. [\onlinecite{arim:07}], the spin-driven multiferroic behavior of CuFeO$_2$ arises from the uniform charge transfer through the metal-ligand hybridization in the presence of spin-orbit coupling. 
Using a combination of INS measurements and theoretical modeling, we have demonstrated that lattice distortions play an important role in determining the magnetic ground state of Ga-doped 
CuFeO$_2$.  With a slight modification of the exchange parameters from pure CuFeO$_2$ and by adjusting 
the anisotropy and distortion energies, we have achieved remarkable agreement
between the observed and predicted dynamical ``fingerprint" of the CNC multiferroic phase.
This complex ground state provides an alternative way to realize multiferroic coupling, where displacements of the oxygen atoms severely distort the spin configuration and produce the electric polarization. Consequently, it is clear that the spin-current model \cite{most:06,kats:05,serg:06} is not the only mechanism responsible for multiferroic behavior and that many other frustrated magnets with rhombohedral or hexagonal symmetries may also exhibit the same form of multiferroic coupling as doped CuFeO$_2$ \cite{arim:10,poie:10}.  The symmetry of these materials makes them candidates for exotic magnetoelectric control like the recently reported magnetic digital flop of ferroelectric domains \cite{seki:09}.

This research was sponsored by the Laboratory Directed
Research and Development Program of Oak Ridge National Laboratory,
managed by UT-Battelle, LLC for the U. S. Department of Energy
under Contract No. DE-AC05-00OR22725 and by the Division of Materials Science
and Engineering and the Division of Scientific User Facilities of the U.S. DOE.


\begin{thebibliography}{PD}

\bibitem{kimu:03} T. Kimura, T. Goto, H. Shintani, K. Ishizaka, T. Arima, and Y. Tokura, Nature (London) {\bf 426}, 55 (2003).

\bibitem{khom:06} D.I. Khomskii, Magn. Magn. Mater {\bf 306}, 1 (2006).

\bibitem{cheo:06}  S.-W. Cheong, and M. Mostovoy, Nature Materials {\bf 6}, 13 (2007).

\bibitem{rame:07} R. Ramesh, and N. A. Spaldin, Nature Materials {\bf 6}, 21 (2007).


\bibitem{Eere:06} W. Eerenstein, N.D. Mathur, and J.F. Scott, Nature {\bf 442}, 759 (2006).

\bibitem{kats:05} H. Katsura, N. Nagaosa, and A. V. Balatsky, Phys. Rev. Lett. {\bf 95}, 057205 (2005).

\bibitem{most:06} M. Mostovoy, Phys. Rev. Lett. {\bf 96}, 067601 (2006).

\bibitem{serg:06}  I. A. Sergienko and E. Dagotto, Phys. Rev. B {\bf 73}, 094434 (2006).

\bibitem{arim:07} T. Arima, J. Phys. Soc. Jpn. {\bf 76}, 073702 (2007).


\bibitem{soda:09} M. Soda, K. Kimura, T. Kimura, M. Matsuura, and K. Hirota, J. Phys. Soc. Jpn. {\bf 78}, 124703 (2009).


\bibitem{seki:07} S.Seki, Y. Yamasaki, Y. Shiomi, S. Iguchi, Y. Onose, and Y. Tokura, Phys. Rev. B {\bf 75}, 100403(R) (2007).

\bibitem{kane:07} S. Kanetsuki, S. Mitsuda, T. Nakajima, D. Anazawa, H. A. Katori, and  K. Prokes, J. Phys.: Condens. Matter {\bf 19}, 145244 (2007).


\bibitem{naka:07} T. Nakajima, S. Mitsuda, S. Kanetsuki, K. Prokes, A. Podlesnyak, H. Kimura, and Y. Noda, J. Phys. Soc. Jpn {\bf 76}, 043709 (2007).

\bibitem{fish:10} R.S. Fishman and S. Okamoto, Phys. Rev. B {\bf 81}, 020402(R) (2010).




\bibitem{ye:07} F. Ye, J.A.  Fernandez-Baca, R.S. Fishman, Y. Ren, H.J. Kang, Y. Qiu, and T. Kimura, Phys. Rev. Lett. {\bf 99}, 157201 (2007).

\bibitem{kimu:06} T. Kimura, J. C. Lashley, and A. P. Ramirez, Phys. Rev. B {\bf 73}, 220401(R) (2006).


\bibitem{tera:05} N. Terada, S. Mitsuda, T. Fujii, K. Soejima, I. Doi, H.A. Katori, and Y. Noda, J. Phys. Soc. Jpn.{\bf 74}, 2604 (2005).

\bibitem{tera:08} N. Terada, T. Nakajima, S. Mitsuda, H. Kitazawa, K. Kaneko, and N. Metoki,  Phys. Rev. B {\bf 78}, 014101 (2008).



\bibitem{fish:08jap} R.S. Fishman, J. App. Phys. {\bf 103} 07B109 (2008).

\bibitem{swan:09} M. Swanson, J.T. Haraldsen, and R.S. Fishman, Phys. Rev. B {\bf 79}, 184413 (2009).


\bibitem{hara:09prl} J.T. Haraldsen, M. Swanson, G. Alvarez, and R.S. Fishman, Phys. Rev. Lett. {\bf 102}, 237204 (2009).



\bibitem{ye:06} F. Ye, Y. Ren, Q. Huang, J.A. Fernandez-Baca, P. Dai, J.W. Lynn, and T. Kimura, Phys. Rev. B {\bf 73} 220404(R) (2006).


\bibitem{tera:06a} N. Terada, S. Mitsuda, H. Ohsumi, and K. Tajima, J. Phys. Soc. Jpn {\bf 75}, 023602 (2006).


\bibitem{fish:08} R.S. Fishman, F. Ye, J.A. Fernandez-Baca, J.T. Haraldsen, and T. Kimura, Phys. Rev. B {\bf 78}, 140407(R) (2008).

\bibitem{hara:09jcp} J. T. Haraldsen and R.S. Fishman, J. Phys.: Condens. Matter {\bf 21}, 216001 (2009).

\bibitem{interactions} The values for the Heisenberg interaction are $J_1$ = -0.19 meV, $J_2$ = -0.099 meV, $J_3$ = -0.13 meV, $J_{z1}$ = -0.049 meV, $J_{z2}$ = 0.019 meV, and $J_{z3}$ = -0.0095 meV. The anisotropy energy is $D$ = 0.0095 meV and distortion energy is $K_1$ = 0.0665 meV.  The resulting harmonics are $C_3 = -8.4\times$10$^{-4}$, $C_5 = -4.0\times$10$^{-5}$, $B_1$ = -0.52, and $B_3$ = -0.011, which are all normalized with respect to the main contribution $C_1$ = 1.
The $C_3$ and $B_3$ harmonics produce excitations at wavevectors [0.4,0.4,1.5] and [0.1,0.1,1.5].  Although these features are too weak to be observed in the 3.5\% Ga-doped compound, their observation in more
weakly doped CuFeO$_2$ would distinguish the CNC phase from the ``proper" spiral in elastic neutron scattering measurements.

\bibitem{senf:07} D. Senff,  P. Link, K. Hradil, A. Hiess, L. P. Regnault, Y. Sidis, N. Aliouane, D. N. Argyriou, and M. Braden, Phys. Rev. Lett. {\bf 98}, 137206 (2007).

\bibitem{arim:10} T. Arima: JPSJ OnlineÑNews and Comments [December 10, 2009].

\bibitem{poie:10} M. Poienar, F. Damay, C. Martin, J. Robert, and S. Petit, Phys. Rev. B {\bf 81}, 104411 (2010).

\bibitem{seki:09} S. Seki, H. Murakawa, Y. Onose, and Y. Tokura, Phys. Rev. Lett. {\bf 103}, 237601 (2009).



\end{thebibliography}
\end{document}